\documentclass[prc,aps,showpacs,twocolumn]{revtex4}
\usepackage{amsmath}
\usepackage{amssymb}
\usepackage{isolatin1}
\usepackage{graphicx}

\begin{document}

\title{Photodisintegration of light nuclei\\ for testing   a  correlated realistic interaction in the continuum}

\author{Sonia Bacca\footnote{electronic address: s.bacca@gsi.de}}

\affiliation{Gesellschaft f\"{u}r Schwerionenforschung, Planckstr.~1,
64291 Darmstadt, Germany}

\date{\today{}}

\begin{abstract}

An exact  calculation of
the photodisintegration cross section of $^3$H, $^3$He and $^4$He is
performed using as interaction the correlated  Argonne V18 potential, constructed
within the  Unitary Correlation Operator Method ($V_{\rm UCOM}$).
 Calculations are carried out using the
 Lorentz Integral Transform method in conjunction with an hyperspherical harmonics
 basis expansion.
A comparison with other realistic potentials and with available
experimental data is discussed.
The $V_{\rm UCOM}$ potential leads to a very similar description of the cross
section as the Argonne V18 interaction with the inclusion of the
Urbana IX three-body force for  photon energies $45\le \omega \le 120$ MeV, while larger differences are found close
to threshold. 
\end{abstract}

\maketitle

\section{introduction}
One of the main challenges in theoretical nuclear physics is
a microscopic description of the properties of finite nuclei,
using  realistic nuclear forces.
For a given interaction model, appropriate observables showing impact on different features  of the nuclear potential need to be tested in a many-body system
through direct comparison with experimental data.
In few-particle systems, where the quantum many-body
problem of nucleons can be solved exactly
both for bound  and scattering states, one has
an optimal setting to probe off-shell properties of different
potentials.

For photodisintegration of light nuclei, exact calculations
predict a  slightly different behavior when using different two-body realistic interactions, 
while a more evident effect  of three-body forces is found.
However, the experimental situation is still unsatisfactory, as the data are
not sufficiently precise to
discriminate clearly among different interactions models.
Among light nuclei,  the photoexcitation of the alpha particle
has recently attracted much attention both in theory,
where a calculation with realistic two- and three-body forces
was carried out \cite{gazit:2006nh},
and in experiments  \cite{nilsson:2005, shima:2005}, where the aim was  to
further clarify whether $^4$He exhibits a pronounced  giant dipole resonance or not, a question that was  raised in the first microscopic calculation of the reaction \cite{efros:1997}, where semirealistic two-body forces where used.
Unfortunately, experimental data  do not yet lead to a unique picture. 
Nevertheless, a comparison of the  impact  on  observables of
different  potentials on a theoretical level is already instructive.

These exact calculations of the photodisintegration cross section can also
offer valuable theoretical guidance in a class of astrophysical studies.
Nuclear absorption of high-energy $\gamma$-rays through excitations
of giant resonances may potentially become an important
diagnostic tool for the dense environments of compact astrophysical sources.
For example, black holes emit an intense high-energy $\gamma$-ray continuum,
against which nuclear absorption features would uniquely probe
the baryonic-matter density, and possibly some compositional information,
near the source.
Since the dependency  on the  potential model used in our calculations turns
out to be
small in comparison to the discrepancy in experiments,
our precise microscopic reaction calculations can help to investigate
interstellar gas, mostly isotopes  of hydrogen
and helium, surrounding the astrophysical $\gamma$-ray source
\cite{iyudin:2005}.

The purpose of this paper is to 
investigate the total photodisintegration cross section of light
nuclei using the potential constructed  within  the
Unitary Correlation Method (UCOM), denoted with $V_{\rm UCOM}$.
This is aimed at
testing for the first time this interaction in a continuum reaction,
where many disintegration channels  are open, and simultaneously
studying the capability of describing electromagnetic reactions
 via  non-local
two-body interactions {\it versus}  local two-and three-body models. 
Exact calculations are performed  making use of the Lorentz
Integral Transform (LIT) approach~\cite{ELO:1994}, which enables to
fully take into account the final state interaction, while 
circumventing the  difficulties arising from the continuum many-body
scattering states. The problem
is in fact reduced to the solution of a bound-state-like equation, which we solve
with an hyperspherical harmonics (HH) basis.

The paper is organized as follows.
In Sec.~\ref{Sec:Overview} the theoretical background is set:
an overview of the UCOM is given, then  the  LIT method is briefly summarized.
In Sec.~\ref{Sec:Results} results are shown
and  conclusions are finally drawn  in Sec.~\ref{Sec:Conclusions}.

\section{Theoretical overview}
\label{Sec:Overview}

\subsection{The Unitary Correlation Operator Method}

Different models for the nucleon-nucleon (NN) interaction 
are found in the literature, as {\it e.g.} the  Argonne V18 potential (AV18) \cite{wiringa:1995}, the CD Bonn \cite{machleidt:2001} and the
Nijmegen \cite{stoks:1994} potentials.
These modern two-body interactions
reproduce the experimental NN data  with high precision, but they 
underbind nuclei with $A\ge3$. This drawback has traditionally been overcome
introducing phenomenological three-body forces fitted to reproduce experimental binding energies and first excited states of light nuclei. 
Furthermore, chiral perturbation theory can provide a systematic way to build two-, three- and more-body forces \cite{entem:2003, epelbaum:2002}.
The above mentioned potential models, including two- and three-body forces have been used to calculate bound state properties of light nuclei (up to $A=12$ and more), mainly within the  Green's function Monte Carlo (GFMC) method 
\cite{pieper:2001, pieper:2002, pieper:2004} and the  no-core shell
model (NCSM) approach \cite{navratil:2000, navratil:2003,
  navratil:2004}.
However, the addition of a three nucleon potential  requires a very intensive computational effort when used in many-body systems.
In Ref.~\cite{polyzou:1990} it was shown that
different, but phase equivalent two-body interactions are related by a
 unitary non-local transformation. It is  expected that  the
 addition of non-locality to the NN interaction could reduce or even
 cancel the need of three-body force, leading to a big simplification
 of many-body calculations.
 Based on this idea, a new category of  NN potentials has then
 emerged, among them one can recall: (i) the Inside Non-local Outside
 Yukawa (INOY) interaction by Doleshall {\em et al.}
 \cite{doleshall:2003},  (ii) the J-matrix Inverse Scattering Potential
 (JISP) by Shirokov {\em et al.} \cite{shirokov:2004, shirokov:2005} and (iii) the Unitary
 Correlation Operator Method (UCOM) by Feldmeier {\em et al.}
 \cite{feldmeier:1998, neff:2003, roth:2004}.
 The $V_{\rm UCOM}$ potential derived from  the AV18 interaction has already been used as universal input in quite a variety
of many-body techniques, ranging form Hartree-Fock calculations
\cite{roth:2006}, to Random Phase Approximation \cite{paar:2006} and  to nuclear structure calculations in the framework of
Fermionic Molecular Dynamics, see {\it e.g.} \cite{neff:2004, neff:2005}.

In the UCOM framework the dominant short range-correlations,  induced  by
the repulsive core and by the tensor part of the NN interaction, are
explicitly described by a state independent unitary transformation. 
When the correlation operator is applied to the Hamiltonian, a phase-shift
equivalent correlated interaction is obtained.
A correlated operator is defined via a
similarity transformation
\begin{equation}
\tilde{O}=\hat{C}^{-1}_{\Omega}\hat{C}^{-1}_{r}\hat{O}\hat{C}_{r}\hat{C}_{\Omega}
\,,
\label{correlated_op}
\end{equation}
where $\hat{C}_{r}$ and $\hat{C}_{\Omega}$  are  the unitary radial
and tensor correlation operators, respectively.
The correlator $\hat{C}_r$  introduces a radial distance-dependent shift
which keeps nucleons away from each other  when their uncorrelated
distance is smaller than the range of the strongly repulsive core of
the NN interaction. $\hat{C}_{\Omega}$ produces tensor correlations by
 further spatial shifts perpendicular to the
radial direction, depending on the orientation of the spins of the two
nucleons with respect to their distance.

In a many-body system the correlated operator of Eq.~(\ref{correlated_op})
becomes an A-body operator with irreducible $n$-body  contributions, $n=1,\dots,A $, 
\begin{equation}
\tilde{O}=\tilde{O}^{[1]}+\tilde{O}^{[2]}+ \tilde{O}^{[3]} +\dots  \tilde{O}^{[A]}\,,
\label{cluster}
\end{equation}
where  $\tilde{O}^{[n]}$ indicates the irreducible $n$-body part. In
the UCOM one usually employs a two-body cluster approximation,
neglecting $ \tilde{O}^{[3]}$ and higher cluster orders.
The effect of this approximation was tested in the framework of NCSM,
where one can perform exact few-body calculations \cite{roth:2005}.
In particular, in this reference  it was shown that the omitted three-
and more-body terms of the cluster expansion can be tuned changing the range of
the tensor correlators in order to compensate the genuine three-body
force to a large extent, still preserving the phase shift equivalence. 
The value of  the tensor  ``correlation volume''
$I_{\theta}^{(1,0)}=0.09$ fm$^3$ (for details see \cite{roth:2005})
was found to give the best description of binding energies of $^3$H
and $^4$He on the Tjon line.  
In this paper we will use this very same value of $I_{\theta}^{(1,0)}$ 
in order to investigate 
whether the non-local two-body $V_{\rm UCOM}$ interaction, that
minimizes the effect of three-body forces on the binding energies, can
also  describe the continuum photoabsorption cross section.

As already pointed out in \cite{polyzou:1990},
phenomenological
 non-local interaction terms introduce modifications on
 the electromagnetic current operator, conserved by gauge invariance.
In fact,
if the nuclear potential $V$ does not commute with the charge
 operator,
then  two-body currents, usually  called Meson Exchange Currents (MEC), have to be introduced, for consistency.
In case of phenomenological non-local potentials the explicit
construction of a consistent MEC could be rather
 involved.
The study of electromagnetic reactions at low energies, where one can use  the Siegert theorem (see {\it e.g.} \cite{Eisenberg-Greiner}), 
allows to investigate the role of implicit
 electromagnetic exchange mechanisms without requiring an explicit knowledge of the two-body current operator.
For this reason, the test  of a given potential model on  the prediction for photoabsorption cross section is very important.

\subsection{The Lorentz Integral Transform approach}

The total photoabsorption cross section is given by
\begin{equation}
\label{0}\sigma(\omega)=4\pi^2 \alpha \omega R(\omega)\,,
\end{equation}
where $\alpha$ is the electromagnetic coupling constant and, at low photon energy  $\omega$, $R(\omega)$ is the
 inclusive unpolarized dipole response function, generally defined as:
\begin{equation}
{\label{1}R(\omega)\!=\!\!\frac{1}{2J_0+1}\!\!\sum_{M_0}\!\! \int \!\!\!\!\!\!\! \sum_f \!\!\left|\left\langle \Psi_{f}\right|\hat{D}_z\left|\Psi_{0}\right\rangle \right|^{2}\!\!\!\delta(E_{f}-E_{0}-\omega)}\,.
\end{equation}
Here, $J_0$ and $M_0$ indicate the total angular momentum of the nucleus in its initial ground state and its projection,  while $\left|\Psi_{0/f}\right>$ and $E_{0/f}$ denote wave function and energies of the ground and final states, respectively. 
The dipole operator is 
\begin{equation}
\hat{D}_z=\frac{1}{2}\sum_i \hat{z}_i \hat{\tau}^3_i \,,
\end{equation}
where $\hat{z}_i$ and $\hat{\tau}^3_i$ are the third components of the position  in the center of mass reference frame and isospin of the $i$-th particle, respectively.
The dipole approximation has been proven to be very good  at low photon energy for the deuteron \cite{arenhoevel:1991} and for the triton case \cite{golak:2002}.
With the dipole operator the major part of the meson exchange currents (MEC) is
implicitly taken into account, via the Siegert theorem.

In the LIT method \cite{ELO:1994}
one obtains $R(\omega)$ after the inversion of an integral transform
with a Lorentzian kernel 
\begin{equation}
L(\sigma_{R},\sigma_{I})=\int d\omega\frac{R(\omega)}{(\omega-\sigma_{R})^{2}+\sigma_{I}^{2}}=\langle \widetilde{\Psi}|\widetilde{\Psi}\rangle \,,\label{2}
\end{equation}
 where $|\widetilde{\Psi}\rangle$ is the unique
solution of the inhomogeneous ``Schr\"{o}dinger-like'' equation
\begin{equation}
(\hat{H}-E_{0}-\sigma_{R}+i\sigma_{I})|\widetilde{\Psi}\rangle=\hat{D}_z|{\Psi_{0}}\rangle.
\label{3}
\end{equation}
Because of the presence of an imaginary part $\sigma_I$ in Eq.~(\ref{3}) 
and the fact that the right-hand side of this  equation is a localized
state, one has an asymptotic
boundary condition similar to a bound state. Thus, one can apply bound-state
techniques for its solution. The response function $R(\omega)$ is  recovered via an inversion of the integral transform of Eq.(\ref{2}) (for inversion methods see Refs. \cite{efros:inv} and \cite{andreasi:05epj}).
Up to now this
method  has allowed  calculations of electromagnetic
reactions  for $^3$H and $^3$He \cite{golak:2002,
  efros:2000, efros:2004, efros:2005} and recently even
for $^4$He \cite{gazit:2006nh} with  realistic two- and three-body
forces.
Nuclei with  six- \cite{Bacca:2001kr,bacca:057001} and  even  seven
nucleons ~\cite{Bacca:2004dr} have been investigated with semirealistic interactions. 
A similar formalism has been  applied to describe
exclusive electromagnetic processes in the four-body system with
semirealistic potentials \cite{quaglioni:044002,quaglioni:064002,andreasi:06eps}. 
Recently, the method has been also applied to the harmonic oscillator
basis in the framework of NCSM \cite{stetcu:2006}.

In the present paper we calculate the LIT using
an HH expansion of the internal wave
function. For the antisymmetrization of the wave function we make use of the
powerful algorithm developed by Barnea {\it et al.}
\cite{barnea:1997,barnea:1998, barnea:1999}.
The HH approach, which  is natural in coordinate space for a local
interaction,  has been recently extended to the use of non-local
potentials given in terms of two-body matrix elements between harmonic
oscillator (HO) states \cite{barnea:2006},  like the JISP potential.
In this paper we will make use of the same method
using as input the correlated HO matrix elements
of the AV18 potential, constructed within the UCOM.

\section{Results}

\begin{figure}[t]
\begin{center}
\rotatebox{0}{\includegraphics*[scale=0.45]{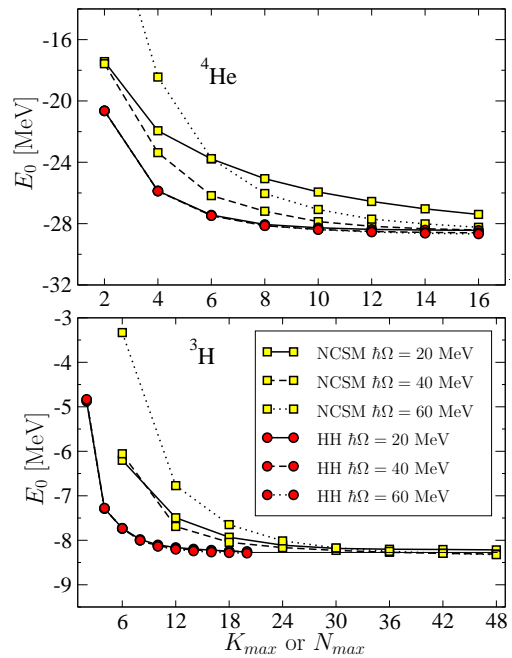}}
\caption{(Color online) Ground-state energies of $^4$He and $^3$H with
  the $V_{\rm UCOM}$ potential for different values of the HO parameter $\hbar\Omega$ 
as a function of the HO excitations $N_{max}$ allowed in the NCSM model space and of the maximal value of the HH grand-angular momentum quantum 
number $K_{max}$ in the HH expansion ($n_{max}=15$).} 
\label{figure1}
\end{center}
\end{figure}
\label{Sec:Results}
We begin the discussion with the results for the ground state energies
of $^4$He and $^3$H calculated in the HH expansion with $V_{\rm UCOM}$.
Even if the starting point is the local AV18 potential, the correlated $V_{\rm UCOM}$ contains derivative operators. In principle, one could extend the  HH formalism to the use of such terms and consider the rather involved operator form of the  $V_{\rm UCOM}$, the result being a more intensive numerical computation for each matrix element. The alternative approach we use, consists in representing the two-body potential operator on an HO basis, similarly to the case of the JISP potential. 
 The input of our HH calculation are the correlated relative two-body HO matrix elements,  
\begin{equation}
v_{nn'\ell \ell' s s'}^{jm, \hbar\Omega}=\left\langle n (\ell s)jm
  |\tilde{v}_{\rm UCOM}| n' (\ell's')jm \right\rangle \!\!\big|_{\hbar\Omega} \,,
\end{equation} 
 where  $n$,$\ell,s$ and $j$ are   radial quantum number,
  orbital angular momentum, spin and 
total angular momentum of the two-body sub-system, respectively (isospin is omitted
for the sake of simplicity).
In the operator form the UCOM is independent on the HO parameters, therefore 
the 
HO representation should  be viewed as a  parameterization rather than a formulation  of the potential. In fact, we expand the two-body potential as
\begin{equation}
\label{potential}
\hat{V}=\!\!
\sum_{nn'}^{n_{max}}
\sum_{ \ell \ell'}^{\ell_{max}}
\!\!\! \sum_{ jmss'}
|n(\ell s)jm \rangle  v_{nn'\ell \ell' s s'}^{jm,\hbar\Omega}
  \langle  n'\! (\ell'\! s'\!)jm| \,.  
\end{equation}
The sum over two-body quantum numbers
$n, n'$ and $\ell, \ell'$ has to be performed up to maximal
value of $n_{max}$ and $\ell_{max}$  for the radial quantum number and
angular momentum, respectively. 
The expansion has to be pushed forward, till 
independence of observables, like binding energies, on the HO parameters is reached.
We find out that good convergence of the angular part is reached with
$\ell_{max}=6$, while  at least $n_{max}=15$
oscillator quanta are needed for the radial part.

Since we work in an
hyperspherical harmonics many-body  Hilbert space, 
convergence as a function of the hyperspherical grand angular momentum
$K_{max}$ needs to be further investigated.
\begin{table}
\caption{Ground state energies in MeV with the $V_{\rm UCOM}$
  potential. NCSM results from \cite{roth:2005}. HH calculations performed with $n_{max}$=15.}
\begin{ruledtabular}
\begin{tabular}{ccc}
Method&{$^4$He}&{$^3$H}\\
\hline
NCSM& -28.4(1) &-8.32(3)\\
HH & -28.57(3) &-8.27(2)\\
Nature & -28.30 &-8.48\\
\end{tabular}
\end{ruledtabular}
\label{table1}
\end{table}
In Fig.~\ref{figure1}, we compare the HH results {\it versus} $K_{max}$ with the
NCSM data from \cite{roth:2005} as a function of  the HO excitation
allowed  $N_{max}=\sum_i 2n_{i}+\ell_i$, where $i$ runs over the $A-1$ Jacobi
coordinates.
Results for different
values of the HO parameter $\hbar \Omega$ are presented. 
 As one can note, the NCSM
 $\hbar \Omega$-dependence is rather strong, while  the HH approach is
 almost HO parameter free. The small residual
 $\hbar\Omega$-dependence, of the order of $0.7$\%, is due to
 the truncation in Eq.~(\ref{potential}). 
The HH convergence as a function of  $K_{max}$ is much
 faster than the NCSM convergence in $N_{max}$. This
 is strongly manifested in case of the more extended nucleus of $^3$H, where convergence of the
 NCSM is reached only with $N_{max}=48$. The reason for that lies in
 the fact
 that  in  the HH-approach  matrix elements with oscillator quanta up to $n_{max}=15$ of the two-body sub-system
  are considered for each $K_{max}$ value, and not only beyond
 $N_{max}=30$, as for the NCSM.
In Table~\ref{table1}  the result for ground state
energies of $V_{\rm UCOM}$ for $^4$He and $^3$H are summarized. The
NCSM results  correspond to the  $\hbar\Omega$ that
yields the minimal energy  for the largest model space used,
whereas for the HH results the mean value of the different
$\hbar\Omega$  has been taken. The two methods agree with each
other. The slightly lower  $^4$He  and the
slightly higher  $^3$H   binding energies in HH are related to the different Hilbert
spaces. They are consistent with the fact that in the HH-approach
with $\ell_{max}=6$ and $n_{max}=15$ one takes into account higher two-body  
HO excitations than with $N_{max}=16$, but less than with $N_{max}=48$.

\begin{figure}
\rotatebox{0}{\includegraphics*[scale=0.47]{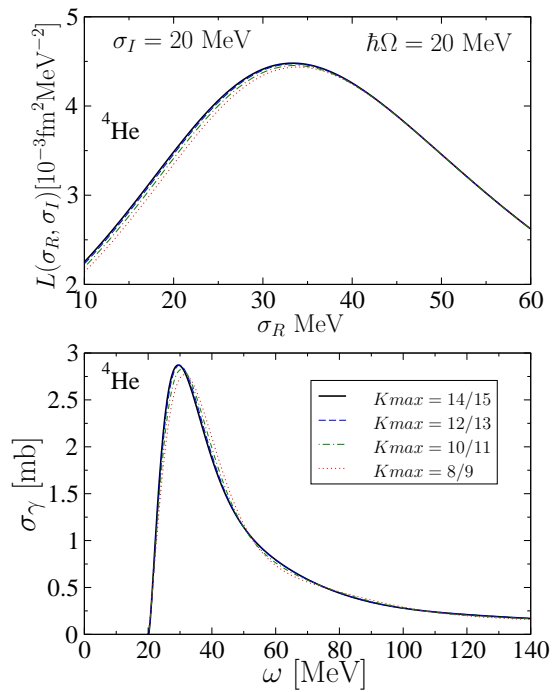}}
\caption{(Color online) $^4$He: (upper panel) convergence of the LIT,
  with fixed $\sigma_I=20$ MeV and HO parameter $\hbar \Omega=20$ MeV,
  for different values  of the hyperspherical grand-angular momentum
  quantum number $K_{max}$ in the HH expansion ($n_{max}=15$); (lower panel) convergence of the photoabsorption cross section.} 
\label{figure2}
\end{figure}

The ground state wave function 
$\left|\Psi_0\right\rangle$ and energy $E_0$
are then used in Eq.~(\ref{3}) to address the problem of the
continuum with the LIT method. 
In Fig.~\ref{figure2}, we show the transforms $L(\sigma_R,
\sigma_I=20~{\rm MeV})$ in case of $^4$He as a function of the parameter
$\sigma_R$ for $\hbar\Omega$ fixed to  $20$ MeV.
The convergence in terms of the HH expansion is studied as a
function of the $K_{max}$: the even/odd value
of $K_{max}$ is due to parity difference of the expanded states,
$|\Psi_{0}\rangle$ and
$\hat{D}|\Psi_{0}\rangle$, respectively.
After inverting  the various transform and making use of Eq.~(\ref{0}) one
can observe the convergence of the photoabsorption cross section.
Full convergence is reached with $K_{max}=14/15$ both
for the LIT and for $\sigma_{\gamma}$. Similar behavior in $K_{max}$,
though slightly weaker,  is obtained for the  $\hbar\Omega$ values of
$40$ and $60$ MeV.
In an analogous way we have  investigated the 
$^3$H case, where  full convergence is reached with  $K_{max}=16/17$. 
The broader structure of $^3$H with respect to  $^4$He makes the rate of
convergence slightly slower.

\begin{figure}
\rotatebox{0}{\includegraphics*[scale=0.57]{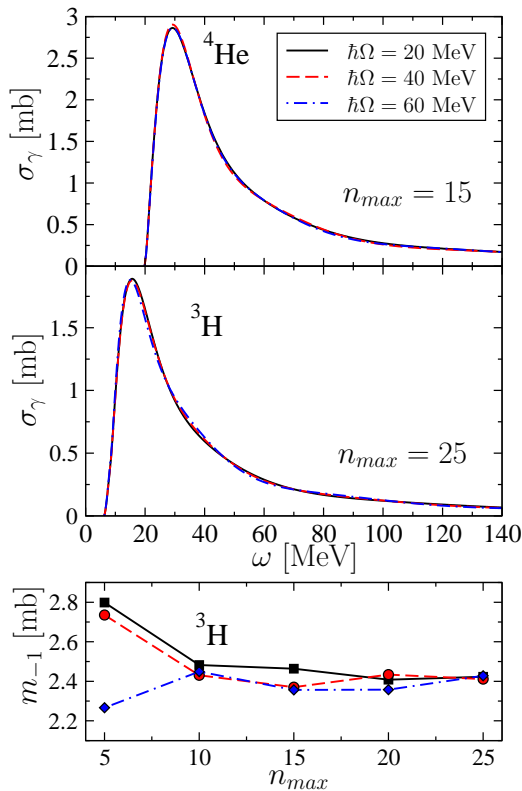}}
\caption{(Color online) Photoabsorption cross section of $^4$He
  (upper panel) and $^3$H (middle panel) for different values of the
  HO parameter $\hbar \Omega$. For $^4$He a maximal value $n_{max}=15$
  in the correlated HO matrix element is used in input, while for
  $^3$H  $n_{max}=25$ is considered. (Lower panel) convergence of the sum rule $m_{-1}$ for $^3$H as a function of $n_{max}$.}
\label{figure3}
\end{figure}

At this point it is convenient to compare the converged cross sections
obtained with different HO parameters to investigate the
$\hbar\Omega$ dependence in the continuum.
As one can see in Fig.~\ref{figure3} in case of $^4$He one gets stable
results for $\sigma_{\gamma}$ with $n_{max}=15$. The 
residual $\hbar\Omega$ dependence can be interpreted as the numerical error
of our calculations: the indetermination is maximally $1.5\%$ in the
dipole resonance peak
region and never exceeds $6\%$ for  higher energies.
On the contrary, in case of  $^3$H the HO parameter dependence is still 
rather strong with $n_{max}=15$, thus higher values of $n
_{max}$ need to be considered in the expansion of Eq.~(\ref{potential}). 
In particular, we observed that for the larger $\hbar\Omega=40$ and
$60$ MeV values,  higher  $n_{max}$ are needed to obtain stable
result for the cross section. 
To check the degree of convergence in the continuum  one can investigate  
the first sum rules of the cross section.
For example, we have considered the inverse energy-weighted sum rule of the cross section $m_{-1}$
(proportional to the total dipole strength),
evaluated as expectation value
on the 
ground state \cite{lipparini},
\begin{equation}
m_{-1}=\!\!\int_{\omega_{th}}^{\infty}
\!\!\frac{\sigma_{\gamma}(\omega)}{\omega}~d\omega=
\!\frac{4\pi^2 \alpha}{3}\left[Z^{2}\left<r_{p}^{2}\right>-
\frac{Z(Z-1)}{2}\left<r_{pp}^{2}\right>\right].
\label{oddo}
\end{equation}
Here  $\langle r_{p}^{2}\rangle$ and $\langle r_{pp}^{2}\rangle$ are
the mean square point-proton 
and mean square proton-proton radii.  
In Fig.~\ref{figure3}, the behavior of the $m_{-1}$ for $^3$H as a function of $n_{max}$
(for converged expansion in $K_{max}$)  is shown  for different HO frequencies.
One can see that for $n_{max}=15$ the discrepancy is of about $5\%$ ($1-2\%$ in case of $^4$He),
while only with  $n_{max}=25$  a satisfactory
$\hbar\Omega$-independence ($0.5\%$) is achieved.
This is again due to the fact that the $^3$H wave function has a longer range structure than $^4$He, thus convergence is slower. 
Furthermore, the higher sensitivity  of $m_{-1}$  in $n_{max}$ with
respect to $E_0$ ($^3$H binding energies decreases of only $30$ KeV going
from $n_{max}=15$ to $25$) is related to the long range nature of the operator.
In the peak region, the final $^3$H photoabsorption cross 
section presents  a $1\%$ difference between the
$\hbar\Omega=20$ MeV and the $\hbar\Omega=40$ MeV results,  and about
$4\%$  between the
$\hbar\Omega=20$ MeV and the $\hbar\Omega=60$ MeV. The reason of the poorer
convergence for a high HO parameter in the continuum 
was already pointed out in \cite{stetcu:2006}, where the LIT method was
applied for the first time to the NCSM basis. Here, though we use an
HH basis, the parameterization of the potential via HO two-body matrix
element leaves some fingerprints, {\it i.e.}
 a slight residual $\hbar\Omega$ dependence. 
As the HO potential well becomes steeper and steeper, the two-body wave functions 
in a fixed model space (fixed $n_{max}$) go faster to zero,
consequence being a poorer representation of 
  long-range operators, like the dipole.
  Thus, higher $n_{max}$ need to be considered to get the
  same degree of of accuracy.
On the contrary, for smaller HO frequencies, the two-body HO
eigenstates are closer to each other, resulting in a better
sampling of the complex-energy continuum of the LIT, and thus in a
better convergence of the cross section.

\begin{figure}
\rotatebox{0}{\includegraphics*[scale=0.45]{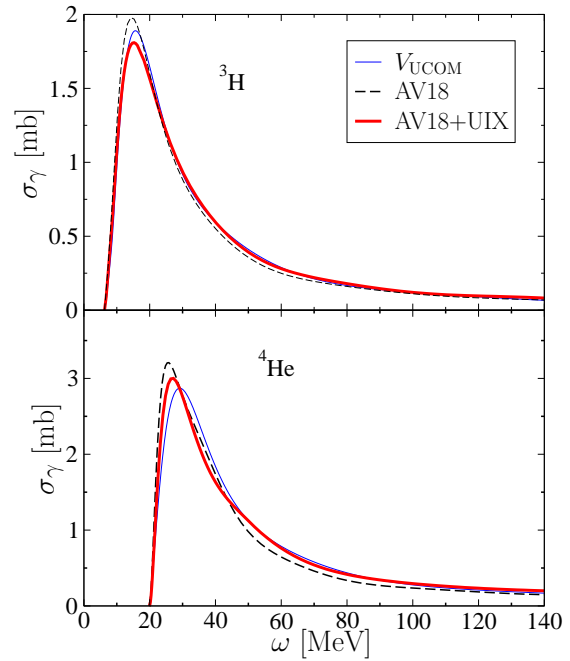}}
\caption{(Color online) Photodisintegration cross section  of triton (upper panel) and $^4$He (lower panel) with different potential models: (thin line) $V_{\rm UCOM}$, (dashed line) bare AV18 and (thick line) AV18+UIX force. 
} 
\label{figure4}
\end{figure}

In Fig.~\ref{figure4}, we compare the photoabsorption cross section
 with $V_{\rm UCOM}$ to those obtained with other potential models: 
 the AV18 potential  
and  the AV18 with the inclusion of the Urbana IX  three-body force, AV18+UIX. 
Predictions for $^3$H and  $^4$He are taken from \cite{golak:2002} and
\cite{gazit:2006nh}, respectively.
The addition of the short range non-locality introduced by the
correlators in the UCOM has consequences on
$\sigma_{\gamma}$. In fact, with respect to the usual AV18 potential,
the $V_{\rm UCOM}$ leads to a reduction of the peak cross section of
about $4\%$ for $^3$H and  $10\%$ for $^4$He, while a general enhancement of
the tail of the cross section is found: it amounts, for example, to  
 $15\%$ and  $23\%$ at $\omega=60$ MeV,  for $^3$H and  $^4$He, respectively.
Similar effects on the photoabsorption cross section are obtained with
the  introduction of the UIX three-body force, though with some differences.
With respect to the interaction model of AV18+UIX the $V_{\rm UCOM}$ potential leads to a $4\%$ higher photodisintegration peak  in case
of $^3$H and to a $4\%$ lower one for $^4$He. We also
observe a $0.5$  and $2$ MeV shift of the peak position towards higher energies
for $^3$H and $^4$He, respectively. At low photon energies prediction
of $V_{\rm UCOM}$ are lower than AV18+UIX: for example the difference
is about $10\%$ for $^3$H and $25\%$ for $^4$He at $3.5$ MeV
after disintegration threshold. 
Interestingly, the tail of the photoabsorption cross section obtained
with the $V_{\rm UCOM}$ is very similar (less than $6-7\%$ difference) to the AV18+UIX result for
photon energy $20\le\omega\le120$ MeV in case of $^3$H and  $45\le\omega\le100$ MeV for the alpha
particle.

\begin{figure}
\includegraphics*[scale=0.47]{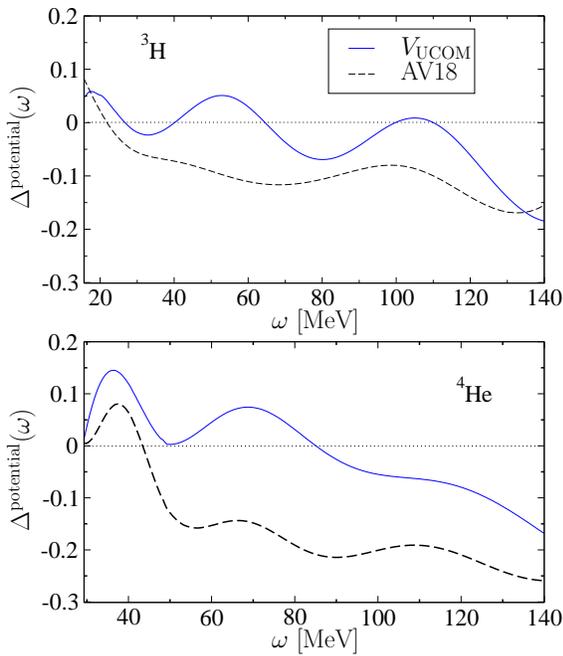}
\caption{(Color online) 
Quantity $\Delta^{\rm potential}(\omega)$ (see text) as a function of
the photon energy in case of $^3$H and $^4$He for the $V_{\rm UCOM}$ and AV18 potentials. } 
\label{figure5}
\end{figure}

We observe that with the  $V_{\rm
  UCOM}$ a  reduction of the missing genuine three-body force is
achieved for high photon energies. In fact, keeping as reference the AV18+UIX curve, the relative
difference on $\sigma_{\gamma}$, defined as
\begin{equation}
\Delta^{\rm potential}(\omega)=\left[1-\frac{\sigma^{\rm
 potential}_{\gamma}(\omega)}{\sigma^{\rm ~AV18+UIX~}_{\gamma}\!\!\!(\omega)} \right]
\end{equation}
gives information on that. In Fig.~\ref{figure5}, the quantity
$\Delta^{\rm potential}(\omega)$ is presented in case of $^3$H and
$^4$He  for $\omega$ beyond the corresponding peak energies obtained
with $V_{\rm UCOM}$. 
In case of $^3$H, $\Delta^{\rm potential}(\omega)$ is reduced  by at least a factor 2 in the energy
region $25\le\omega\le75$ MeV and  $90\le\omega\le120$, going from the AV18 to the  $V_{\rm UCOM}$ potential.
An even stronger reduction effect is found for $^4$He for energies between $45$ MeV and pion threshold.

\begin{table}
\begin{ruledtabular}
\begin{tabular}{cccc}
Potential& $\left\langle r^2_p\right\rangle$ $^3$H &$\left\langle
  r^2\right\rangle$ $^4$He  & $\left\langle r_{pp}^2\right\rangle$  $^4$He  \\
\hline
$V_{\rm UCOM}$&2.52(1) &1.96(1) &5.41(1)\\

AV18+UIX &2.51(1) & 2.05(1)&5.67(1)\\
\end{tabular}
\end{ruledtabular}
\caption{Mean square  point proton and proton-proton radii  in fm$^2$
  with $V_{\rm UCOM}$ and AV18+UIX ($\left\langle r^2_p\right\rangle =\left\langle
    r^2\right\rangle$ for $^4$He without isospin-mixing).
  Results for triton from \cite{gazit:private} and for alpha particle
  from \cite{gazit:2006}.}
\label{table2}
\end{table}

In the following an investigation of two photonuclear sum rules, the already mentioned $m_{-1}$ (Eq.~(10)) and the  Thomas-Reiche-Kuhn (TRK) sum rule \cite{threku}
\begin{eqnarray}
m_{0}&=&\int_{\omega_{th}}^{\infty} \sigma_{\gamma}(\omega)~d\omega\\
\nonumber
& =&  \frac{4\pi^2 \alpha}{2J_0+1} \sum_{M_0} \left<\Psi_0; M_0
    |[\hat{D}_z,[\hat{H},\hat{D}_z]]|\Psi_0; M_0\right>\,,
\end{eqnarray}
 is  briefly presented (see also Ref.~\cite{gazit:2006}).
With respect to the AV18+UIX case, the $V_{\rm UCOM}$ potential predicts a  very similar $m_{-1}$ for $^3$H and a lower  $m_{-1}$ for $^4$He. This fact  is also reflected in the size of the radii of  Eq.(\ref{oddo}).
In Table~\ref{table2} we summarize the situation:  for $^3$H we present only the 
 proton radius, since the there is no proton-proton radius, while  for $^4$He we show the squared mass radius
($\left\langle r^2 \right\rangle=\left\langle r^2_{p}\right\rangle$ if isospin mixing is neglected) and $\left\langle r^2_{pp}
\right\rangle$.
One can see that $V_{\rm UCOM}$ leads almost to the same proton radius as AV18+UIX within a $0.4\%$ deviation, but 
 smaller radii for the alpha particle are obtained, if compared to  AV18+UIX.
A consequence of that  is  the lower $^4$He peak cross section  found with $V_{\rm UCOM}$  with respect to AV18+UIX.
The TRK sum rule is of interest since it
contains information on   the exchange mechanisms induced by the
non-commuting part of the potential $\hat{V}$.
For example, for $^4$He, we find that $m_{0}=117.6$ mb MeV for $V_{\rm UCOM}$,
which is about $20\%$  lower in comparison to the value of $146.2$ mb
MeV found for AV18+UIX \cite{gazit:2006}. 
This  is more strongly
connected to the fact that with $V_{\rm UCOM}$ the cross section goes
faster to zero beyond pion threshold and indicates that the meson exchanges underlying the two potential models are different.

\begin{figure}
\includegraphics*[scale=0.60]{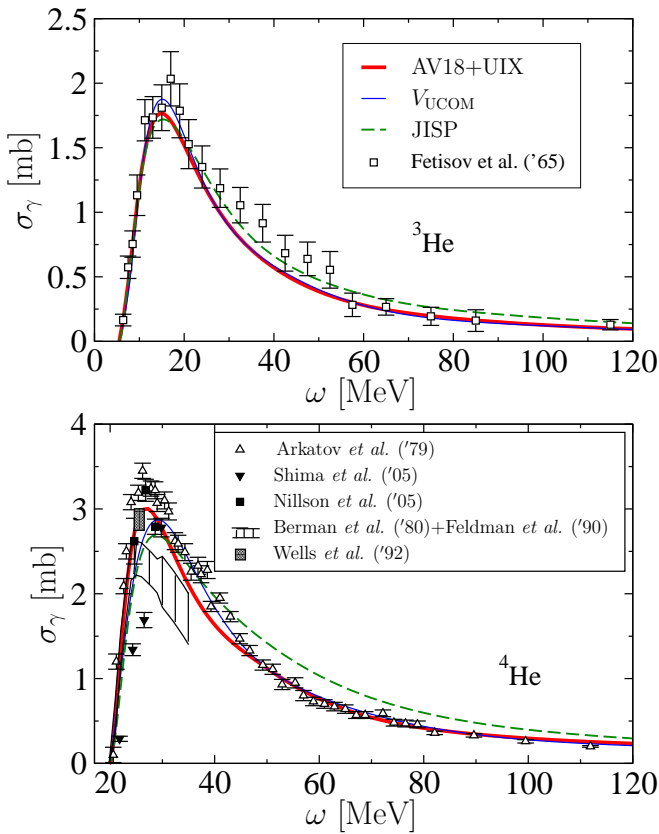}
\caption{(Color online) 
Theoretical photoabsorption cross section of $^3$He and $^4$He with
three different potential models, (thin line) $V_{\rm UCOM}$, (thick
line) AV18+UIX and (dashed line) JISP potential from
\cite{barnea:2006}, in comparison with the available experimental
data: empty squares from \cite{fetisov:1965} for $^3$He, empty  triangles from
\cite{arkatov:1979}, full triangles from \cite{shima:2005}, full squares
from  \cite{nilsson:2005}, shaded area sum of data from \cite{berman:1980} and \cite{feldman:1990} and box from \cite{wells:1992} for $^4$He. 
 } 
\label{figure6}
\end{figure}

We now  compare the theoretical predictions with the
available experimental data. 
Among the theoretical curves  we  add the result with the
JISP potential \cite{shirokov:2005}, recently published in
\cite{barnea:2006}. 
In Fig.~\ref{figure6}, we show the situation for the $^3$He nucleus and for $^4$He.
In case of the three-body nucleus the error bars of the  data from
Fetisov {\it et al.} \cite{fetisov:1965} are unfortunately too big to allow us to
discriminate among the  different  potential models.

In case of the alpha particle the experimental situation is more
involved.
Close to threshold several data were taken in different experiments,
which unfortunately show fairly large discrepancies. Only the data from
Arkatov {\it et al.} 
\cite{arkatov:1979} cover a larger energy range. They present a rather
high peak cross section, in  fair agreement with recent data from
\cite{nilsson:2005},  which favors the AV18+UIX
potential model.
In the energy region $30\le\omega\le 45$ MeV the results with $V_{\rm UCOM}$ agree better
 with this set of data  than those with AV18+UIX.
Finally, at higher energies the measurements are precise enough to 
conclude that  AV18+UIX and $V_{\rm UCOM}$ can explain, within the
error bars, 
the photoabsorption cross section, in contradistinction to the JISP potential.
In Ref.~\cite{barnea:2006} it was argued that a possible reason of
this discrepancy of the  JISP model could be due to the probably
incorrect long range part of the potential, whose construction  is not
constrained by the
meson exchange theory. Regarding this issue we can only state that
 $V_{\rm UCOM}$, which explicitly contains the long-range pion exchange term
of  AV18, unchanged by the short range action of the correlations,
exhibits a correct tail behavior in the photoabsorption cross section
of light nuclei.
Here we would also like to point out that the consideration of large
Hilbert spaces (large $K_{max}$ and $n_{max}$) is essential to
achieve a stable and HO-independent result on the whole photon energy
range. The fact that the JISP potential is constructed on a small HO basis,
for  defined  $n_{max}$ and $\hbar\Omega$ values, could limit the
correct description of the tail of this sensitive observable.

\section{Conclusions}
\label{Sec:Conclusions}

We have presented the results of an \emph{ab initio} calculation 
of the $^3$H, $^3$He and $^4$He photodisintegration cross section with
the $V_{\rm UCOM}$ potential. The difficulty of the scattering
continuum problem is circumvented via the Lorentz Integral Transform  method, where the
final state interaction is fully taken into account. An expansion
of the wave function on hyperspherical harmonics, extended for non-local
interaction, is used for the solution of the ``Schr\"{o}dinger-like'' equation.
The first investigation of the Unitary Correlation Operator Method  on a
continuum observable is presented in this paper. The sensitivity of the
photoabsorption  cross section on non-local terms of the interaction
is investigated. With respect to the traditional local AV18 potential, a reduction of the peak and an
enhancement of the tail of the cross section is produced by the
non-locality of $V_{\rm UCOM}$.
The comparison between the AV18+UIX
potential model and $V_{\rm UCOM}$  allows to investigate, whether
 the omission of a genuine three-body force can be replaced by a
 non-local interaction on a continuum observable.
Though  binding energies of very light nuclei are well
described by $V_{\rm UCOM}$ with a  $1-2\%$ difference with respect
to the AV18+UIX model, the situation is different in the photodisintegration cross section below pion
threshold.
Our analysis shows that the $V_{\rm UCOM}$ potential leads to a very similar
result as the AV18+UIX for high photon energy,
 while in the region close to threshold 
larger differences  are found, particularly in case of  $^4$He. 
The larger deviation in $^4$He
with respect to the $^3$H case could be related
to the fact that smaller  mean square radii are obtained with the
$V_{\rm UCOM}$. 
 The higher density of the alpha particle makes it certainly more
sensitive to characteristic  short range properties of the three-body
force and non-local parts of the NN interaction, magnifying the
differences. 
The similarity of the high-energy  cross
section shows that also in the continuum  genuine three-body
forces can be partly simulated with the non-local  $V_{\rm UCOM}$ potential.
But the differences found at low-photon energies point out that the two
potential models are not completely equivalent.
Unfortunately, the lack of precise experimental data limits the
possibility to use  such calculations as a discriminant test of the
potentials.

\begin{acknowledgments}
The author would like to thank Nir Barnea for providing the HH code extended
for the use of non-local interaction, and Robert Roth  for supplying the
correlated harmonic oscillator matrix element of $V_{\rm UCOM}$.
Further gratefulness is owed to
Hans Feldmeier, Winfried Leidemann  and Giuseppina Orlandini for many
useful discussions and for a critical reading of the manuscript.
Numerical calculations are partly performed at CINECA (Bologna).
 
\end{acknowledgments}



\end{document}